\documentclass[twocolumn,showpacs,superscriptaddress,preprintnumbers,amsmath,amssymb,prc]{revtex4-2}

\usepackage{graphicx}
\usepackage{dcolumn}
\usepackage{bm}
\usepackage{float}
\usepackage{color}
\usepackage{ulem}
\usepackage{CJK}
\usepackage[colorlinks,linkcolor=blue,urlcolor=blue,citecolor=blue]{hyperref}

\usepackage{tikz,xcolor,hyperref}

\definecolor{lime}{HTML}{A6CE39}
\DeclareRobustCommand{\orcidicon}{
	\begin{tikzpicture}
	\draw[lime, fill=lime] (0,0) 
	circle [radius=0.16] 
	node[white] {{\fontfamily{qag}\selectfont \tiny ID}};
	\draw[white, fill=white] (-0.0625,0.095) 
	circle [radius=0.007];
	\end{tikzpicture}
	\hspace{-2mm}
}
\foreach \x in {A, ..., Z}{%
	\expandafter\xdef\csname orcid\x\endcsname{\noexpand\href{https://orcid.org/\csname orcidauthor\x\endcsname}{\noexpand\orcidicon}}
}

\begin{document}
\begin{CJK*}{UTF8}{gbsn}

\title{Impact of magnetic field on giant dipole resonance of $^{40}$Ca using the EQMD model}
\author{Y. T. Cao(曹雅婷)}
\affiliation{Key Laboratory of Nuclear Physics and Ion-beam Application (MOE), Institute of Modern Physics, Fudan University, Shanghai 200433, China}

\author{X. G. Deng(邓先概)\orcidB{} \footnote{ xiangai\_deng@fudan.edu.cn}}
\affiliation{Key Laboratory of Nuclear Physics and Ion-beam Application (MOE), Institute of Modern Physics, Fudan University, Shanghai 200433, China}
\affiliation{Shanghai Research Center for Theoretical Nuclear Physics， NSFC and Fudan University, Shanghai 200438, China}

\author{Y. G. Ma(马余刚)\orcidC{} \footnote{mayugang@fudan.edu.cn}}
\affiliation{Key Laboratory of Nuclear Physics and Ion-beam Application (MOE), Institute of Modern Physics, Fudan University, Shanghai 200433, China}
\affiliation{Shanghai Research Center for Theoretical Nuclear Physics， NSFC and Fudan University, Shanghai 200438, China}

\date{\today}

\begin{abstract}

By taking into account the magnetic field in the extended quantum molecular dynamics model (EQMD), we analyzed its effects on giant dipole resonance (GDR)  by studying the responses and strengths of the dipole moments. The selected system is the $^{40}$Ca nucleus which is excited through the Coulomb interaction by $^{16}$O.  Particle acceleration term in Li\'enard-Wiechert potential is discussed which, however, has small impact on magnetic field. The peak energy, strength and width of GDR, temperature, and angular momentum of $^{40}$Ca as a function of beam energy are investigated. It is found that the magnetic field enhances the peak energy, strength and width of GDR which is  not only due to the temperature effects but also due to the enhancement of the angular momentum of nucleus.  At beam energy {E} $>$ 200 MeV/nucleon, magnetic field maintains a constant value for the strength of GDR. The work sheds light  on examining important roles of the magnetic field on nuclear structure in low-intermediate energy heavy-ion collisions.

\end{abstract}


\maketitle

\section{Introduction}
\label{Introduction}
The electric and magnetic fields generated in heavy-ion collisions as well as  their effects on observable are of research interests for some times~\cite{Rafelski76}. The strongest magnetic field in the universe could be produced  in  ultra-relativistic heavy-ion collisions by theoretical prediction~\cite{Skokov09,Voronyuk11}. Some observational effects and its properties for this strong field have been discussed in relativistic heavy-ion collisions~\cite{Fukushima08,Kharzeev08,Asakawa10,MGL15,HXG17}, but less discussed in low-intermediate energy domain \cite{Ou11,DXG17,Wei18,Sun19,Deng20}. Even though the magnetic field is weaker in this energy region in contrast to ultra-relativistic collisions, but its lifetime is longer, its roles on nuclear structure or reaction should be further surveyed.

Giant resonances (GR) are resonance states built in the nucleus~\cite{Harakeh01}, which are of particular importance because they  provide reliable information about the bulk behavior of nuclear many-body systems~\cite{Rep,Ait21}. One of the best-known modes of giant resonances is isovector giant dipole resonance (IVGDR) in which neutrons and protons move collectively relative to each other in a nucleus. IVGDR is an excellent tool for systematical investigation on collective properties of nuclei because it can give crucial clues to understand nuclear structure and collective dynamics. Furthermore, IVGDR can also give rise to a dynamic electric-dipole (E1) moment~\cite{JHB21}, which reflects asymmetry information in nuclear equation of state (EoS)~\cite{Bracco19,Trippa08,Pie,Tamii,XuJun}. Experimentally, the IVGDR can be induced by various ways such as photoabsorption~\cite{Carlos71,Carlos74,RMP,Masur06}, 
inelastic scattering~\cite{Donaldson18,Ramakrishnan96} as well as ultra-peripheral Coulomb excitation \cite{Bertulani} etc., and the  data can be fitted using the Lorentzian curve from which the key parameters consisting of the peak energy, the width, and the peak strength can be extracted. The peak energy is a good probe of the symmetry energy $E_{sym}$ around and below the saturation density~\cite{Trippa08}, and it can also provide direct information about nuclear size and the EoS~\cite{Harakeh01,Pie} as well as the shear viscosity \cite{eta1,GuoCQ,eta2,Srijit21}. Meanwhile, it is well established that the width and peak splitting can be taken as a direct experimental tool to measure the nuclear deformation~\cite{PRL1,Pandit13,SSW17,Pandit20}.

Most studies of GRs are built on the ground state of nuclei. In Ref.~\cite{Axel62}, however, it is conjectured that GRs could be built on other nuclear states and their characteristics would not depend on the detailed microscopic structure which gives rise to the enthusiasm to study the properties of GRs built on excited states. In addition, the IVGDR is considered to be one of the most useful tools to study  properties of rotating nuclei at finite temperature~\cite{Alhassid99}. The resonance properties of IVGDR, such as peak energy, width and peak strength, are investigated as a function of temperature, contributing  on understanding of the symmetry energy term in  the nuclear equation of state~\cite{BALi,Liu}.

Several microscopic approaches have been employed to study giant resonances such as separable random-phase-approximation (SRPA)~\cite{Kleinig08,Nesterenko07}, time-dependent Skyrme-Hartree-Fock (TSHF) method~\cite{Maruhn05,Fracasso12}, relativistic quasi-particle random phase approximation (RQRPA)~\cite{Arteaga09,Colo2,PhysRep} and some transport models, for instance, Boltzmann-Uhlenbeck-Uehling (BUU) model \cite{Wang_PLB,Wang_fphy,Song_PRC} and quantum molecular dynamics (QMD) model~\cite{SSW17}. Among them, the extended quantum molecular dynamics (EQMD) model ~\cite{Maruyama96} which was improved based upon the spirit of QMD has been successfully applied to investigate IVGDR ~\cite{WBH14,WBH16}.

A variety of experimental and theoretical studies have been carried out by far, but there are a few discussions about the Coulomb excitation, especially in the excited states. Besides, the impact of magnetic field is seldom investigated at low and intermediate energy heavy-ion reaction which is far from pion production energies~\cite{DXG17}. So the aim of this paper is to study the effect of electromagnetic field on the IVGDR of the excited $^{40}$Ca through ultra-peripheral $^{40}$Ca + $^{16}$O  collision at energy ranging from 50 to 500 MeV/nucleon.

The organization of the paper is as follows: In Sect.~\ref{Model&Method}, we give a brief introduction of our simulation model and method including electromagnetic field and the algorithm  of IVGDR. Results of magnetic field effects are discussed in Sect.~\ref{Results&Discussion}. Finally, conclusion is given in Sect.~\ref{summary}.

\section{Model and methodology}
\label{Model&Method}

\subsection{EQMD model}
\label{EQMD Model}

Quantum molecular dynamics (QMD) model has been widely used to study intermediate energy nuclear reactions and nuclear fragmentation~\cite{Aichelin91}. It has the microscopic basis and high flexibility of this type of models~\cite{Furuta10} that make them successful application in studies of giant resonances such as giant dipole resonance (GDR), pygmy dipole resonance (PDR) as well as giant monopole resonance (GMR)~\cite{Kanada05,HLW10,Huang2021}. Nevertheless, standard QMD model shows insufficient stability, for which the phase space obtained from the samples of Monte Carlo is usually not in the lowest point of energy~\cite{Maruyama96}. To compensate for this shortcoming, two features are of great importance, one is the capability to describe nuclear ground state, another is the stability of nucleus in the model description~\cite{Maruyama96}. The above both features have been well solved in the EQMD model. As we know, in order to cancel the zero-point energy caused by the wave packet broadening in the standard QMD, the friction cooling process can be used to keep the mathematical ground state, but the Pauli principle is broken~\cite{Maruyama96}. However, the EQMD model embodies the Pauli potential into effective interactions, making saturation property and $\alpha$-cluster structures can be obtained after energy cooling~\cite{WBH14}. Besides, the EQMD model takes the kinetic-energy term of the momentum variance of wave packets in the Hamiltonian into account, which is ignored as the spurious constant term in the standard QMD~\cite{Aichelin86,Hartnack98}. Furthermore, in the EQMD model, the Hamiltonian introduces the wave packet width as a complex variable, and treats it as an independent dynamic variable, and in contrast, a uniform and static width is applied for all nucleons in the standard QMD model.

In the EQMD model, the wave packet of nucleon is taken as the form of Gaussian-like, of which the total wave function of the system is treated as a direct product of all nucleons~\cite{Maruyama96}
\begin{equation}
\Psi=\prod_{i}\varphi(\boldsymbol{r}_i)     \,       ,
\label{Wavepacket01}
\end{equation}

\begin{equation}
\varphi(\boldsymbol{r}_i)=(\frac{\nu_{i}+\nu^{\ast}_{i}}{2\pi})^{3/4}{\rm exp}[-\frac{\nu_{i}}{2}(\boldsymbol{r}_i-\boldsymbol{R}_i)^2+\frac{i}{\hbar}\boldsymbol{P}_i\cdot\boldsymbol{r}_i]     \,       ,
\label{Wavepacket02}
\end{equation}
where $\boldsymbol{R}_i$ and $\boldsymbol{P}_i$ are the centers of position and momentum of the $i$-th wave packet, respectively. The Gaussian width $\nu_{i}$ is introduced as

\begin{equation}
\nu_{i}\equiv\frac{1}{\lambda_{i}}+i\delta_{i}     \,       ,
\label{Gaussianwidth}
\end{equation}
where $\lambda_{i}$ and $\delta_{i}$ are dynamical variables in the process of initialization.

The energy-minimum state is considered as the ground state of initial nucleus. At the first step, a random configuration is given to each nucleus. Furthermore, under the time-dependent variation principle (TDVP)~\cite{Kerman76},  propagation of each nucleon can be described as~\cite{Maruyama96}
\begin{equation}
\begin{split}
\boldsymbol{\dot{R}}_{i}=\frac{\partial{H}}{\partial{\boldsymbol{P}_{i}}}+\mu_{\boldsymbol{R}}\frac{\partial{H}}{\partial{\boldsymbol{R}_{i}}},\boldsymbol{\dot{P}}_{i}=-\frac{\partial{H}}{\partial{\boldsymbol{R}_{i}}}+\mu_{\boldsymbol{P}}\frac{\partial{H}}{\partial{\boldsymbol{P}_{i}}},\\
\frac{3\hbar}{4}\dot{\lambda}_{i}=-\frac{\partial{H}}{\partial{\delta_{i}}}+\mu_{\lambda}\frac{\partial{H}}{\partial{\lambda_{i}}},\frac{3\hbar}{4}\dot{\delta}_{i}=\frac{\partial{H}}{\partial{\lambda_{i}}}+\mu_{\delta}\frac{\partial{H}}{\partial{\delta_{i}}},
\label{Dampedequations}
\end{split}
\end{equation}
where $H$ is the expected value of the Hamiltonian, and $\mu_{\boldsymbol{R}}$, $\mu_{\boldsymbol{P}}$, $\mu_{\lambda}$ and $\mu_{\delta}$ are various friction coefficients. During the friction cooling process, these coefficients are negative, making the system goes to its local energy minimum point~\cite{SSW17}. In contrast, in the subsequent nuclear reaction simulation stage, these coefficients are zero to maintain the energy conservation of the system~\cite{CZS21}.

As a consequence, the expected value of Hamiltonian can be expressed as
\begin{equation}
\begin{split}
H& = \left\langle\Psi\left|\sum_{i}-\frac{\hbar^{2}}{2m}\bigtriangledown^{2}_{i}-\hat{T}_{zero}+\hat{H}_{int}\right|\Psi\right\rangle\\
& = \sum_{i}\frac{\boldsymbol{P}^{2}_{i}}{2m}+\frac{3\hbar^{2}(1+\lambda^{2}_{i}\delta^{2}_{i})}{4m\lambda_{i}}-T_{zero}+H_{int},
\label{Hamiltonian}
\end{split}
\end{equation}
where the first, second and third term are the center momentum of the wave packet, the contribution of the dynamic wave packet, and the zero point center-of-mass kinetic energy $-T_{zero}$, respectively. The first term can be expressed as $\left\langle\hat{\boldsymbol{p}}_{i}\right\rangle^{2}/2m$,  the second term can be treated as $\left(\left\langle\hat{\boldsymbol{p}}_{i}^{2}\right\rangle-\left\langle\hat{\boldsymbol{p}}_{i}\right\rangle^{2}\right)/2m$, and the form of the third term can be found in details in Ref.~\cite{Maruyama96}.

For the effective interaction $H_{int}$, it consists of the Skyrme potential, the Coulomb potential, the symmetry energy, and the Pauli potential as follows
\begin{equation}
H_{int} = H_{Skyrme} + H_{Coulomb} + H_{Symmetry} + H_{Pauli}.
\label{Effectiveinteraction}
\end{equation}

The form of Skyrme interaction is the simplest, written as
\begin{equation}
H_{Skyrme} = \frac{\alpha}{2\rho_{0}}\int\rho^{2}(\boldsymbol{r})d^{3}r+\frac{\beta}{(\gamma+1)\rho_{0}^{\gamma}}\int\rho^{\gamma+1}(\boldsymbol{r})d^{3}r,
\label{Skyrmeinteraction}
\end{equation}
where $\alpha = -124.3$ MeV, $\beta = 70.5$ MeV, and $\gamma = 2$, which can be obtained by the fits to the ground state properties of finite nuclei.

A phenomenological repulsive Pauli potential is introduced to prevent nucleons with the same spin-$S$ and isospin-$I$ to come close to each other in the phase space, which can be presented as
\begin{equation}
H_{Pauli} = \frac{c_{P}}{2}\sum_{i}(f_{i}-f_{0})^{\mu}\theta(f_{i}-f_{0}),
\label{Paulipotential01}
\end{equation}
\begin{equation}
f_{i}\equiv\sum_{j}\delta(S_{i},S_{j})\delta(I_{i},I_{j})\left|\left\langle\phi_{i}|\phi_{j}\right\rangle\right|^{2},
\label{Paulipotential02}
\end{equation}
where $f_{i}$ is 
the overlap of the $i$-th nucleon with other nucleons having the same spin and isospin, 
$\theta$ is the unit step function, and $c_{P} = 15$ MeV is a coefficient denoting strength of  the Pauli potential. For the other two parameters, we take $f_{0} = 1.0$ and $\mu = 1.3$. And the symmetry potential can be written as
\begin{equation}
H_{Symmetry} = \frac{C_{S}}{2\rho_{0}}\sum_{i,j\neq{i}}\int[2\delta(I_{i},I_{j})-1]\rho_{i}(\boldsymbol{r})\rho_{j}(\boldsymbol{r})d^{3}r,
\label{Symmetrypotential}
\end{equation}
where $C_{S}$ is the symmetry energy coefficient which is  25 MeV in this work.

\subsection{Li\'enard-Wiechert potential}
\label{Retardedpotential}

Li\'enard and Wiechert proposed that the electromagnetic potential of a moving particle can be obtained from a scalar potential field $\phi$ and a vector potential field $\boldsymbol{A}$ by Lorentz gauge respectively as follows~\cite{Griffiths99}
\begin{equation}
\phi(\boldsymbol{r},t) = \frac{1}{4\pi\epsilon_{0}}\left(\frac{q}{(1-\boldsymbol{n}\cdot\boldsymbol{v}_{s}/c)|\boldsymbol{r}-\boldsymbol{r}_{s}|}\right)_{t_{s}}     \,      ,
\label{Retardedpotential01}
\end{equation}
\begin{equation}
\boldsymbol{A}(\boldsymbol{r},t) = \frac{\mu_{0}}{4\pi}\left(\frac{q\boldsymbol{v}_{s}}{(1-\boldsymbol{n}\cdot\boldsymbol{v}_{s}/c)|\boldsymbol{r}-\boldsymbol{r}_{s}|}\right)_{t_{s}}     \,      ,
\label{Retardedpotential02}
\end{equation}
where $\epsilon_{0}$ and $\mu_{0}$ are the dielectric constant and permeability in vacuum, the speed of light is $1/\sqrt{\epsilon_{0}\mu_{0}}$, and $q$ is the charge of the particle.  $\boldsymbol{n}$ is the unit vector represented by  $(\boldsymbol{r}-\boldsymbol{r}_{s})/|\boldsymbol{r}-\boldsymbol{r}_{s}|$, and $\boldsymbol{v}_{s}$ and $\boldsymbol{r}_{s}$ are the velocity and position of the particle at time $t_{s}$, respectively.
The potential field $\phi(\boldsymbol{r},t)$ and $\boldsymbol{A}(\boldsymbol{r},t)$ at $t = t_{s}+|\boldsymbol{r}-\boldsymbol{r}_{s}|/c$ which is excited by particle $q$ at $t_{s}$ is called the Li\'enard-Wiechert potential and often referred as the retarded potential.

The electromagnetic field generated by moving charged particles can be expressed as
\begin{widetext}
\begin{eqnarray}
e\boldsymbol{E}(\boldsymbol{r},t) &=& \frac{e^{2}}{4\pi\epsilon_{0}}\sum_{n}\frac{Z_{n}}{(cR_{n}-\boldsymbol{R}_{n}\cdot\boldsymbol{v}_{n})^{3}}
\{(c^{2}-v_{n}^{2})(\boldsymbol{R}_{n}-R_{n}\boldsymbol{v}_{n})+(\boldsymbol{R}_{n}\cdot\boldsymbol{a}_{n})(c^{2}\boldsymbol{R}_{n}-cR_{n}\boldsymbol{v}_{n}) \notag \\
&-&[c^{2}R_{n}^{2}-cR_{n}(\boldsymbol{R}_{n}\cdot\boldsymbol{v_{n}})]\boldsymbol{a}_{n}\}     \,      ,\label{Electricfield01}\\
e\boldsymbol{B}(\boldsymbol{r},t) &=& \frac{e^{2}}{4\pi\epsilon_{0}c}\sum_{n}\frac{Z_{n}}{R_{n}(cR_{n}-\boldsymbol{R}_{n}\cdot\boldsymbol{v}_{n})^{3}}
\{R_{n}(c^{2}-v_{n}^{2})(\boldsymbol{v}_{n}\times\boldsymbol{R}_{n})
+ cR_{n}(\boldsymbol{R}_{n}\cdot\boldsymbol{a}_{n})(\boldsymbol{v}_{n}\times\boldsymbol{R}_{n})   \notag\\
&+&\boldsymbol{R}_{n}(c\boldsymbol{R}_{n}-cR_{n}\boldsymbol{v}_{n})(\boldsymbol{a}_{n}\times\boldsymbol{R}_{n})\} \,,
\label{Magneticfield01}
\end{eqnarray}
\end{widetext}
where $Z_{n}$ and $\boldsymbol{a}_{n}$ is the number of charges and acceleration on the particle $n$, respectively;
and $\boldsymbol{R}_{n} = \boldsymbol{r} - \boldsymbol{r}_{sn}$, where $\boldsymbol{r}_{sn}$ and $\boldsymbol{v}_{n}$ are the position and velocity of the particle $n$ at retarded time $t_{sn} = t - |\boldsymbol{r}-\boldsymbol{r}_{sn}(t_{sn})|/c$, respectively. In particular, the charge constant $q$ is multiplied on the Eq.~(\ref{Electricfield}) and the Eq.~(\ref{Magneticfield}), so that the right sides of them contain the fine structure constant $\alpha = e^{2}/(4\pi) = 1/137$ $(\epsilon_{0} = \hbar = c = 1)$.
If we ignore the acceleration of charged particles, the electromagnetic field can be simplified as~\cite{Ou11,Voronyuk11}
\begin{equation}
e\boldsymbol{E}(\boldsymbol{r},t) = \frac{e^{2}}{4\pi\epsilon_{0}}\sum_{n}Z_{n}\frac{c^{2}-v_{n}^{2}}{(cR_{n}-\boldsymbol{R}_{n}\cdot\boldsymbol{v}_{n})^{3}}(c\boldsymbol{R}_{n}-R_{n}\boldsymbol{v}_{n})     \,      ,
\label{Electricfield}
\end{equation}
\begin{equation}
e\boldsymbol{B}(\boldsymbol{r},t) = \frac{e^{2}}{4\pi\epsilon_{0}c}\sum_{n}Z_{n}\frac{c^{2}-v_{n}^{2}}{(cR_{n}-\boldsymbol{R}_{n}\cdot\boldsymbol{v}_{n})^{3}}\boldsymbol{v}_{n}\times\boldsymbol{R}_{n}     \,      ,
\label{Magneticfield}
\end{equation}

In the non-relativistic approximation, that is, $v\ll{c}$, the Eq.~(\ref{Electricfield}) and the Eq.~(\ref{Magneticfield}) can be further simplified as
\begin{equation}
e\boldsymbol{E}(\boldsymbol{r},t) = \frac{e^{2}}{4\pi\epsilon_{0}}\sum_{n}Z_{n}\frac{\boldsymbol{e}_{n}}{R_{n}^{3}}     \,      ,
\label{nonElectricfield}
\end{equation}
\begin{equation}
e\boldsymbol{B}(\boldsymbol{r},t) = \frac{e^{2}}{4\pi\epsilon_{0}c}\sum_{n}Z_{n}\frac{\boldsymbol{v}_{n}\times\boldsymbol{R}_{n}}{R_{n}^{3}}     \,      .
\label{nonMagneticfield}
\end{equation}

For self-consistency, we use the original Coulomb potential of the EQMD model for the electric field part, expressed as
\begin{equation}
H_{Coulomb} = \frac{e^{2}}{2}\sum_{i}\sum_{i\neq{j}}Z_{i}Z_{j}\frac{1}{r_{ij}}{\rm erf}(\frac{r_{ij}}{\sqrt{4L}})     \,      ,
\label{EQMDCoulombpotential01}
\end{equation}
where
\begin{equation}
r_{ij} = |\boldsymbol{r}_{i}-\boldsymbol{r}_{j}| \, ,
\label{EQMDCoulombpotential02}
\end{equation}
\begin{equation}
{\rm erf}(x) = \frac{2}{\sqrt{\pi}}\int^{x}_{0}e^{-u^{2}}du     \,      ,
\label{EQMDCoulombpotential03}
\end{equation}
and the magnetic field can be obtained from the Eq.~(\ref{Magneticfield01}) or Eq.~(\ref{Magneticfield}) for making a comparison.

\begin{figure}[htb]
\setlength{\abovecaptionskip}{0pt}
\setlength{\belowcaptionskip}{0pt}
\centering\includegraphics[scale=0.6]{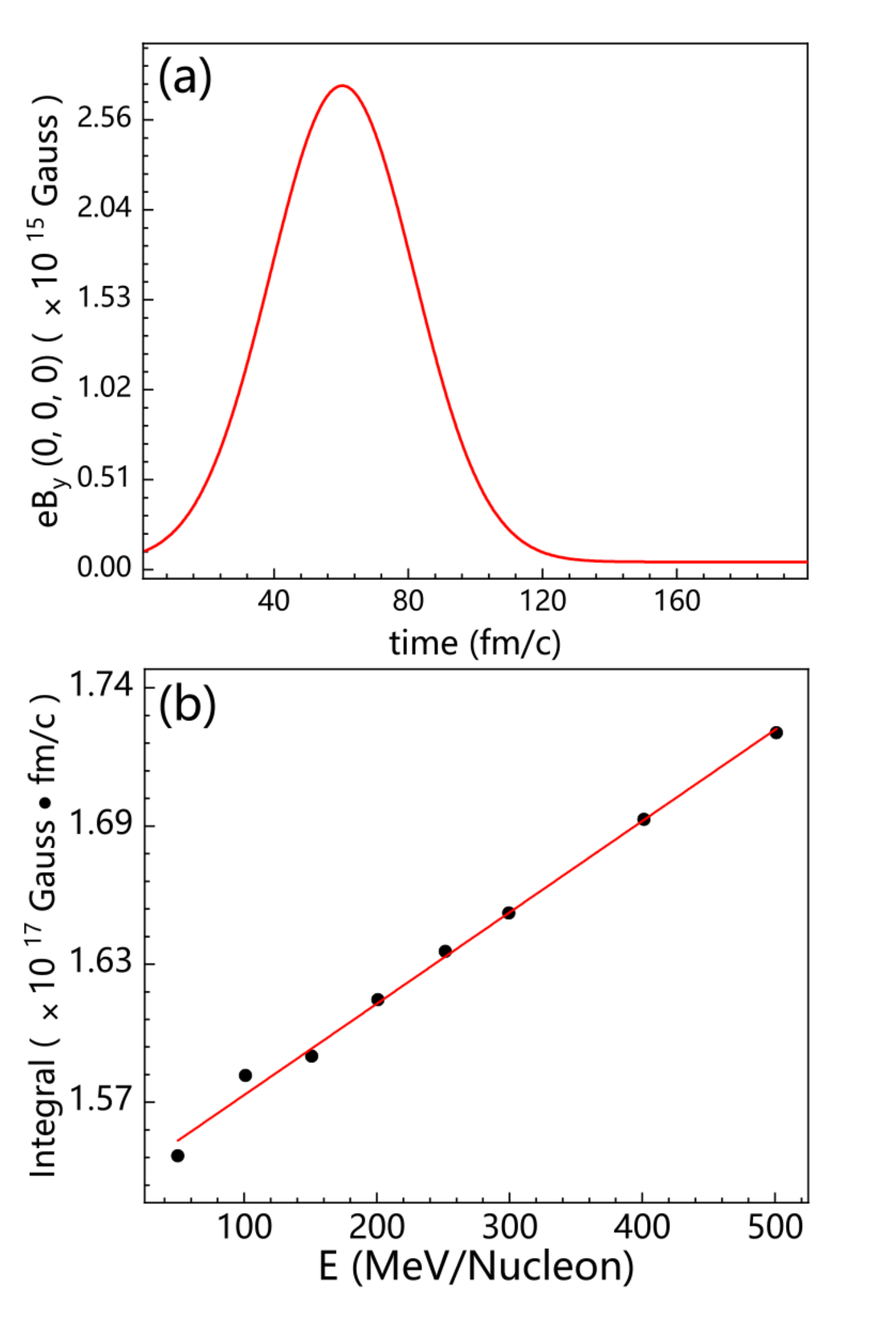}
\caption{Time evolution of the magnetic field at the central point (0,0,0)  of $^{16}$O + $^{40}$Ca at beam energy of 100 MeV/nucleon and impact parameter of 7.13 fm (a). The dependence between magnetic field integral over time and beam energy for $^{16}$O + $^{40}$Ca at 7.13 fm (b).}
\label{fig:B_time}
\end{figure}

With the Eq.~(\ref{Magneticfield}), we displayed time evolution of the magnetic field at different beam energies. For an example, Fig.~\ref{fig:B_time}(a) shows  time evolution of the magnetic field at the central point (0,0,0) produced in  $^{16}$O + $^{40}$Ca collision process at 100 MeV/nucleon and impact parameter of 7.13 fm. One can see that the maximum   strength of the magnetic field reaches $2.75 \times 10^{15}$ Gauss (in Lorentz units, 1 MeV$^{2}\approx$ 5.11 $\times$ 10$^{13}$ Gauss) at $\sim$ 60 $\rm fm/c$ for 100 MeV/nucleon collision and the lifetime of the magnetic field can extend to $\sim$ 120 $\rm fm/c$. By integrating over the time evolution of the magnetic field, we can get  the magnetic field integral as a function of beam energy as shown in Fig.~\ref{fig:B_time}(b), from which we observed an almost linear correlation. 

\subsection{GDR algorithm}
\label{GDR}

Macroscopic description of IVGDR by the Goldhaber-Teller model~\cite{Goldhaber48} is applied to calculate GDR~\cite{Sonia13,Sonia14} from the nuclear phase space obtained by the EQMD model. There are two ways to calculate IVGDR spectrum, the one is to give the nucleus a boost to obtain the dipole oscillation, the another is to simulate a Coulomb excitation with a  nucleus~\cite{WBH16}. In this work, we chose the latter one.

We get the initial state wave function $\Psi(0)$ of nucleus by the EQMD model and let $^{16}$O interact 
with the target nucleus $^{40}$Ca at some energies and an impact parameter $7.13$ fm, which is the sum of radii of $^{40}$Ca and $^{16}$O.
Besides, the process to obtain the evolution of excited wave function to the final state is  described in Ref.~\cite{Maruyama96} in details.

The dipole moments of the system in the coordinate space $D_{G}(t)$ and momentum space $K_{G}(t)$ can be written as~\cite{WBH14,WBH16,SSW17,HLW10,Baran01}
\begin{equation}
D_{G}(t) = \frac{NZ}{A}[R_{Z}(t)-R_{N}(t)]     \,      ,
\label{GDR01}
\end{equation}
\begin{equation}
K_{G}(t) = \frac{NZ}{A\hbar}\left[\frac{P_{Z}(t)}{Z}-\frac{P_{N}(t)}{N}\right]     \,      ,
\label{GDR02}
\end{equation}
where $R_{Z}(t)[P_{Z}(t)]$ and $R_{N}(t)[P_{N}(t)]$ are the center-of-mass of the protons and neutrons in the coordinate (momentum) space, respectively,  $A$ is the sum of mass numbers of the target and projectile nuclei, $N$ and $Z$ are the neutron and proton numbers of the compound system, respectively.  $K_{G}(t)$ is the canonically conjugate momentum of $D_{G}(t)$~\cite{WBH14}.

\begin{figure}[htb]
\setlength{\abovecaptionskip}{0pt}
\setlength{\belowcaptionskip}{0pt}
\centering\includegraphics[scale=0.58]{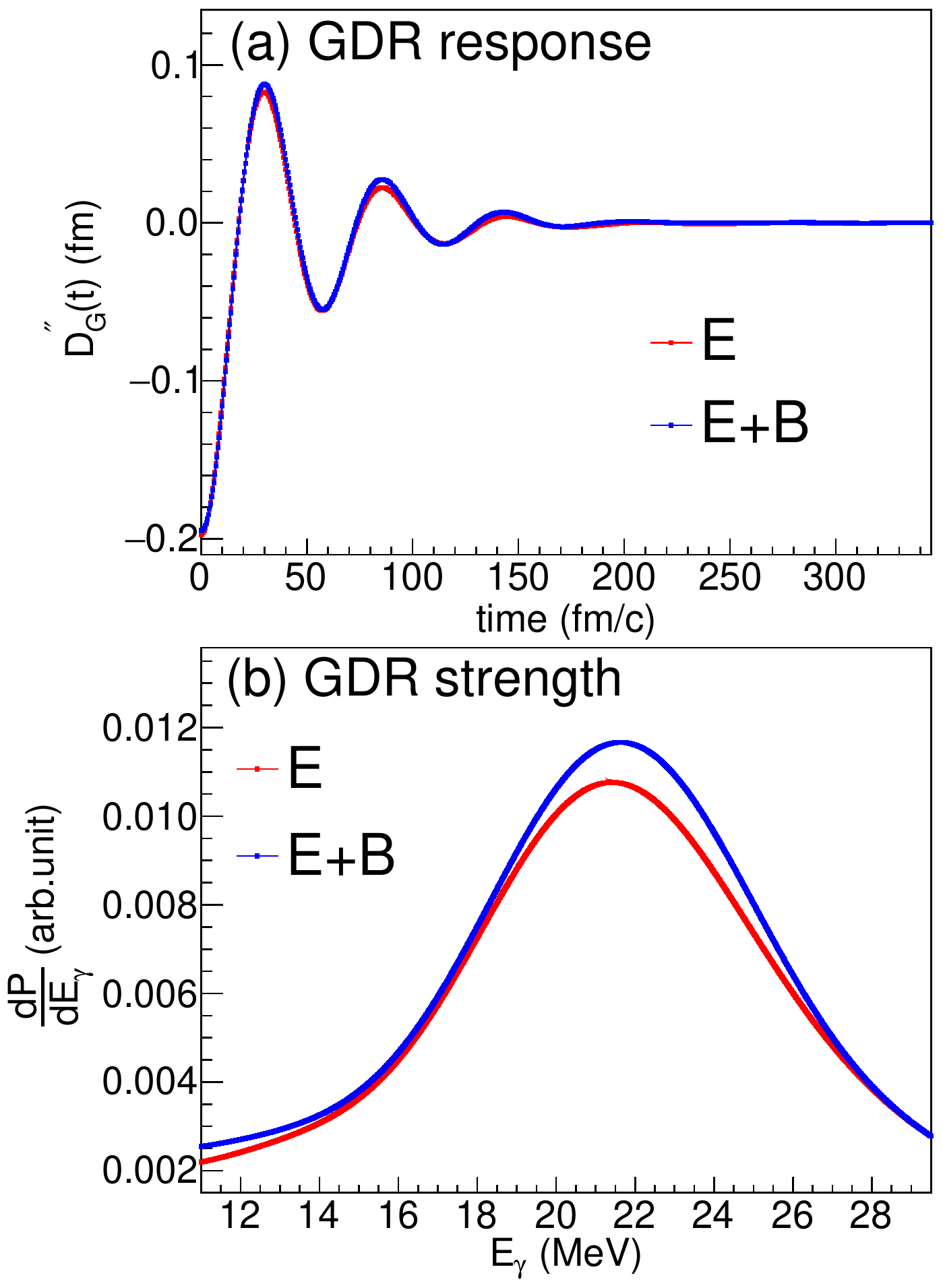}
\caption{Time evolution of the dipole moments (a) and $\gamma$-emission probability (b) for $^{40}$Ca 
excited by $^{16}$O at incident energy $100$ MeV/nucleon.}
\label{fig:GDRMagneticfield}
\end{figure}

The $\gamma$-emission probability of nuclear system at photon energy $E_{\gamma} = \hbar\omega$ can be derived from Eq.~(\ref{GDR01}) as
\begin{equation}
\frac{dP}{dE_{\gamma}} = \frac{2e^{2}}{3\pi\hbar c^{3}E_{\gamma}}|D^{\prime\prime}(\omega)|^{2}     \,      ,
\label{GDR03}
\end{equation}
where $D^{\prime\prime}(\omega)$ is obtained from the Fourier transform of the second derivative of $D_{G}(t)$ with respect to time, i.e.,
\begin{equation}
D^{\prime\prime}(\omega) = \int^{t_{2}}_{t_{1}}D^{\prime\prime}_{G}(t)e^{i\omega t}dt     \,      .
\label{GDR04}
\end{equation}

Considering the lifetime of GDR excitation, we take the final state at $t_{2} = 200 \, \text{fm}/c$, which corresponds with the lifetime of GDR excitation.

\section{Results and discussion}
\label{Results&Discussion}

\begin{figure}[htb]
\setlength{\abovecaptionskip}{0pt}
\setlength{\belowcaptionskip}{0pt}
\centering\includegraphics[scale=0.58]{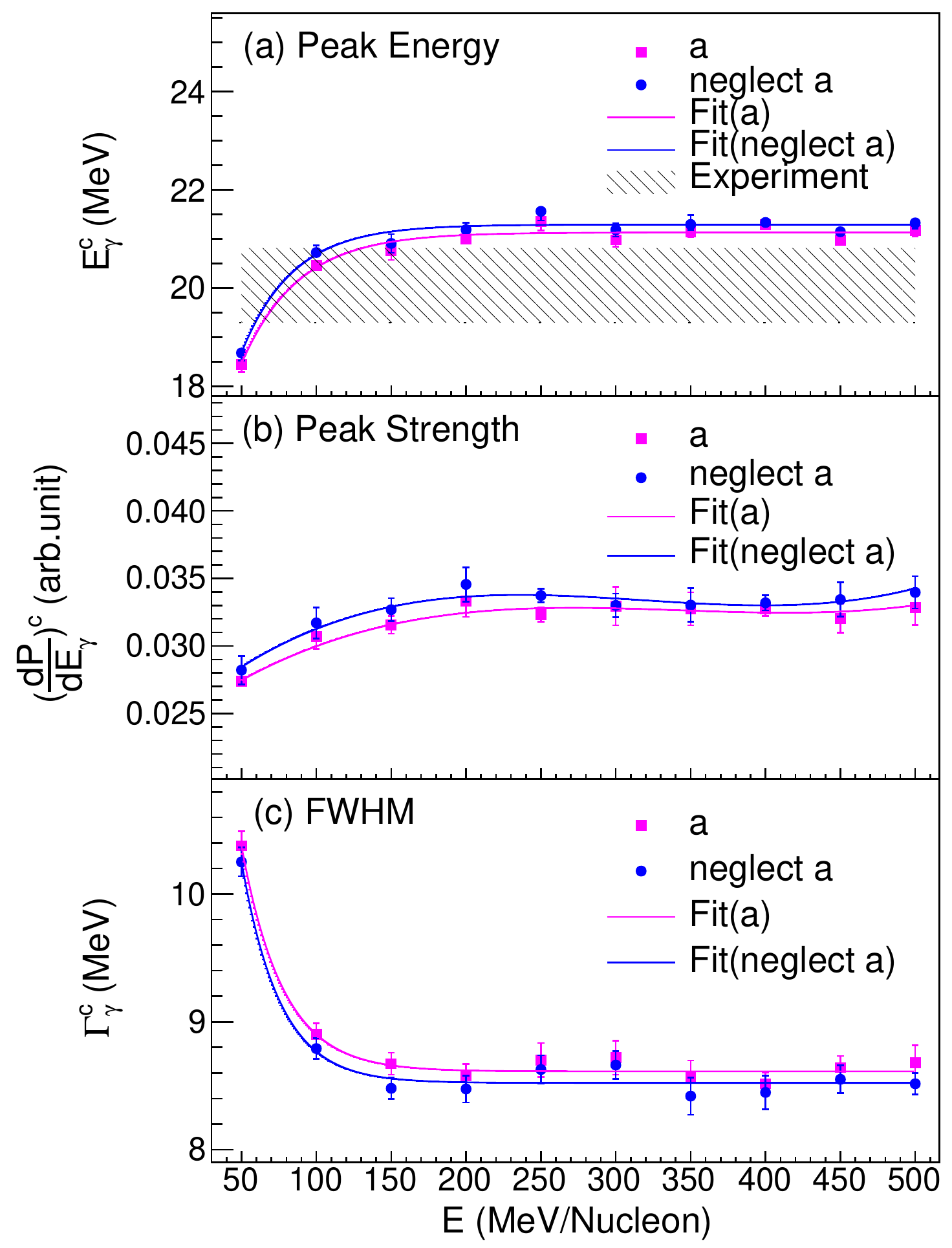}
\caption{The dependence of peak energy (a), strength (b) and width of GDR spectrum (c) on incident energy for $^{40}$Ca excited by $^{16}$O in the cases whether or not to consider acceleration. The symbol denoted ``a" represents the results with particle acceleration effect, while ``neglect a" represents the ones without. The lines are fits to the dots. The hatched area represents the experimental result.}  
\label{fig:ImpactA}
\end{figure}

\begin{figure}[htb]
\setlength{\abovecaptionskip}{0pt}
\setlength{\belowcaptionskip}{0pt}
\centering\includegraphics[scale=0.58]{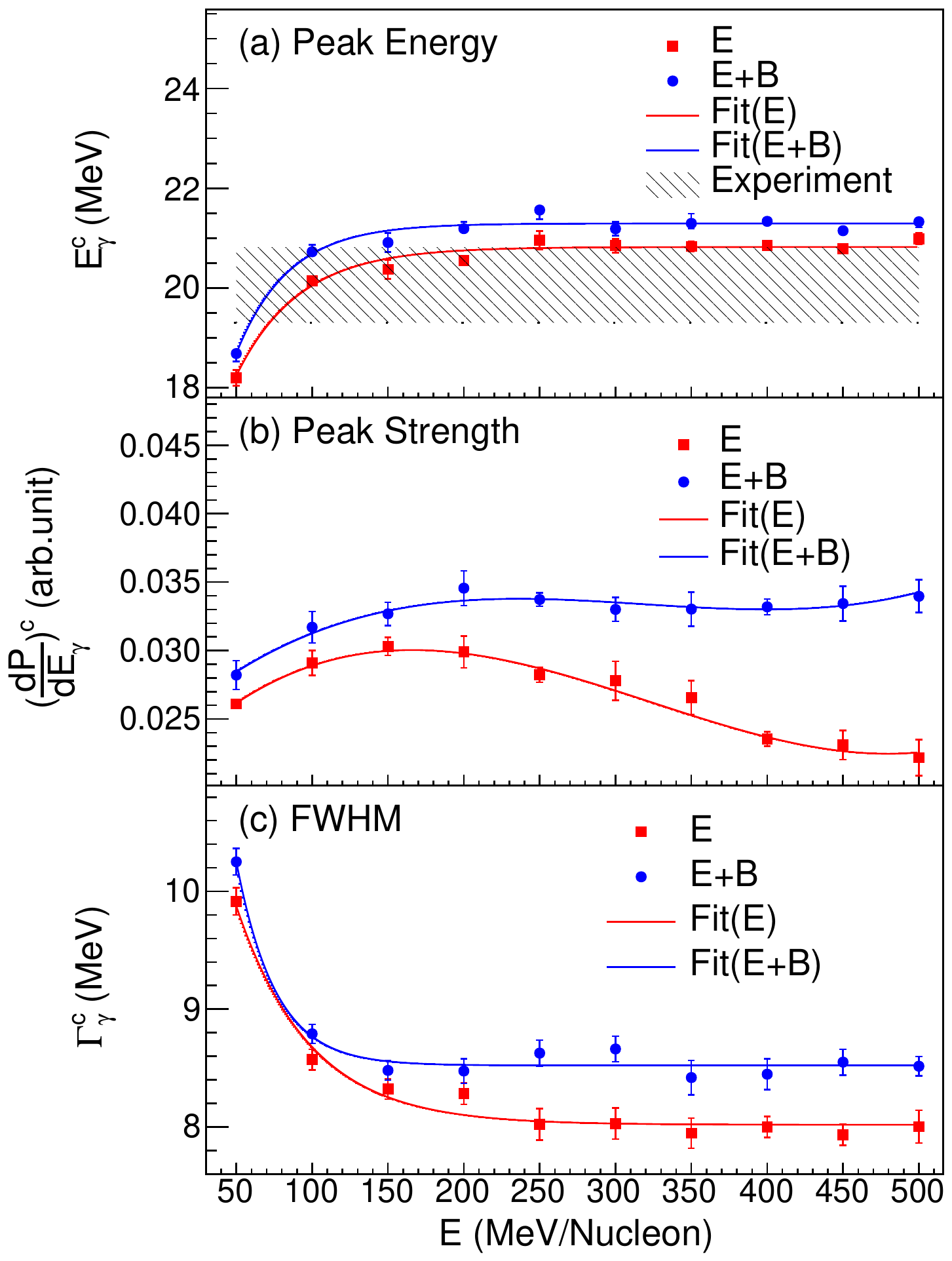}
\caption{The dependence of peak energy (a), strength (b) and width of GDR spectrum (c) on incident energy for $^{40}$Ca excited by $^{16}$O in the cases without or with magnetic field.}
\label{fig:ImpactMagneticfield}
\end{figure}

\subsection{Impacts on GDR of $^{40}$Ca}
\label{ResultsA}

We discuss ultra-peripheral collision case (here it means b = $7.13$ fm) of $^{40}$Ca and $^{16}$O, and focus on the GDR analysis for excited $^{40}$Ca. Many previous studies~\cite{WBH14,WBH16} have shown the existence of $\alpha$-cluster structure in $^{16}$O. For simplicity,  we only consider $^{16}$O whose nucleons are distributed by the Woods-Saxon distribution without any $\alpha$-clusters. The electromagnetic fields come from the $^{40}$Ca and $^{16}$O interactions. In our simulations, more than 90\% of $^{40}$Ca remain complete, and the following results have excluded the events that $^{40}$Ca undergoes fragmentation.  First of all, we check the dipole moment $D_{G}(t)$ and $\gamma$-emission probability for $^{40}$Ca with definitions in Eqs.~(\ref{GDR01}) and (\ref{GDR03}) which are shown in Fig.~\ref{fig:GDRMagneticfield}. Seeing from Fig.~\ref{fig:GDRMagneticfield} (a), dipole moment $D_{G}(t)$ seems not to be affected by the magnetic field. However, one can see that $\gamma$-emission probability is slightly affected by the magnetic field and the peak value increases a little bit.

\begin{figure}[htb]
\setlength{\abovecaptionskip}{0pt}
\setlength{\belowcaptionskip}{0pt}
\centering\includegraphics[scale=0.45]{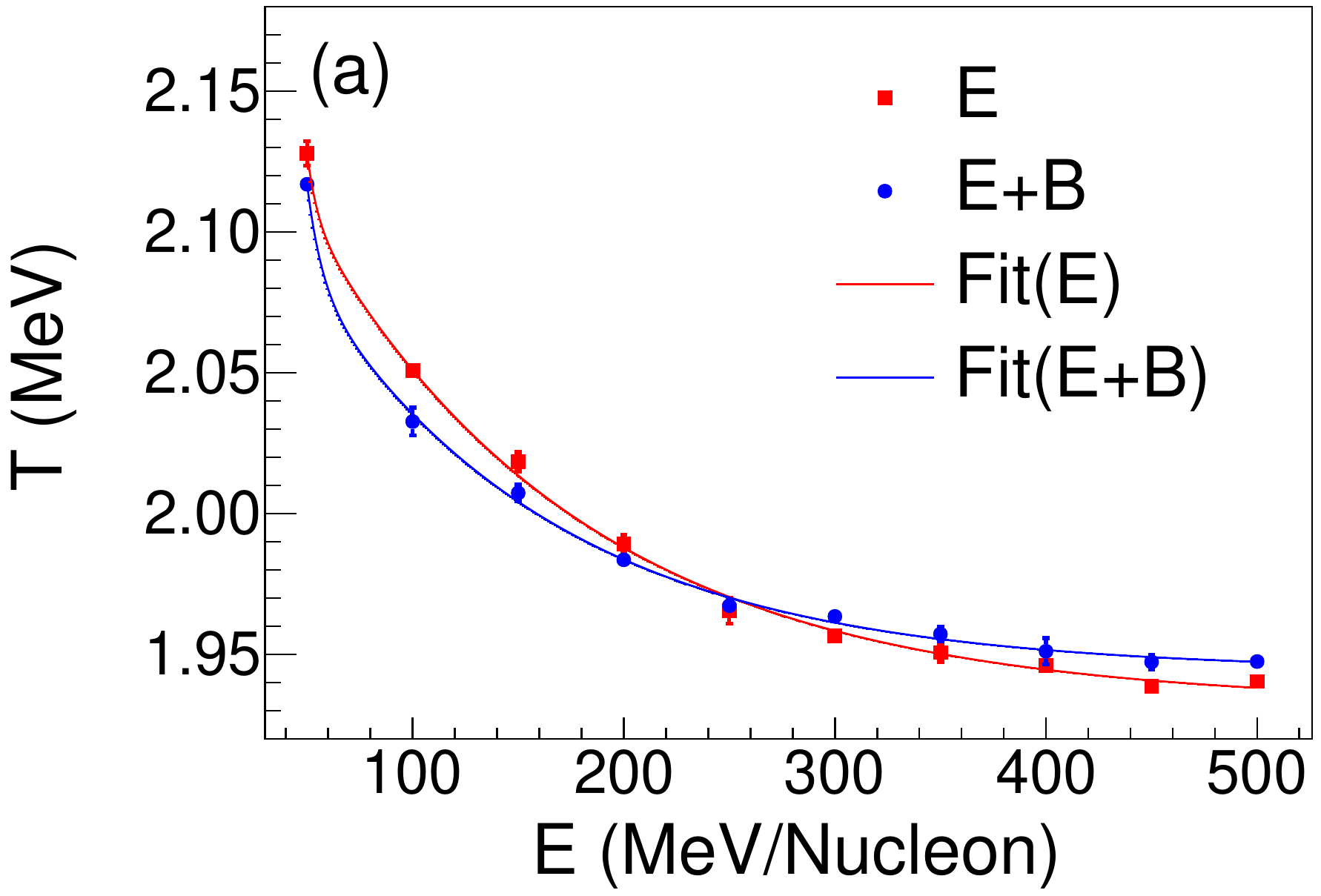}
\centering\includegraphics[scale=0.45]{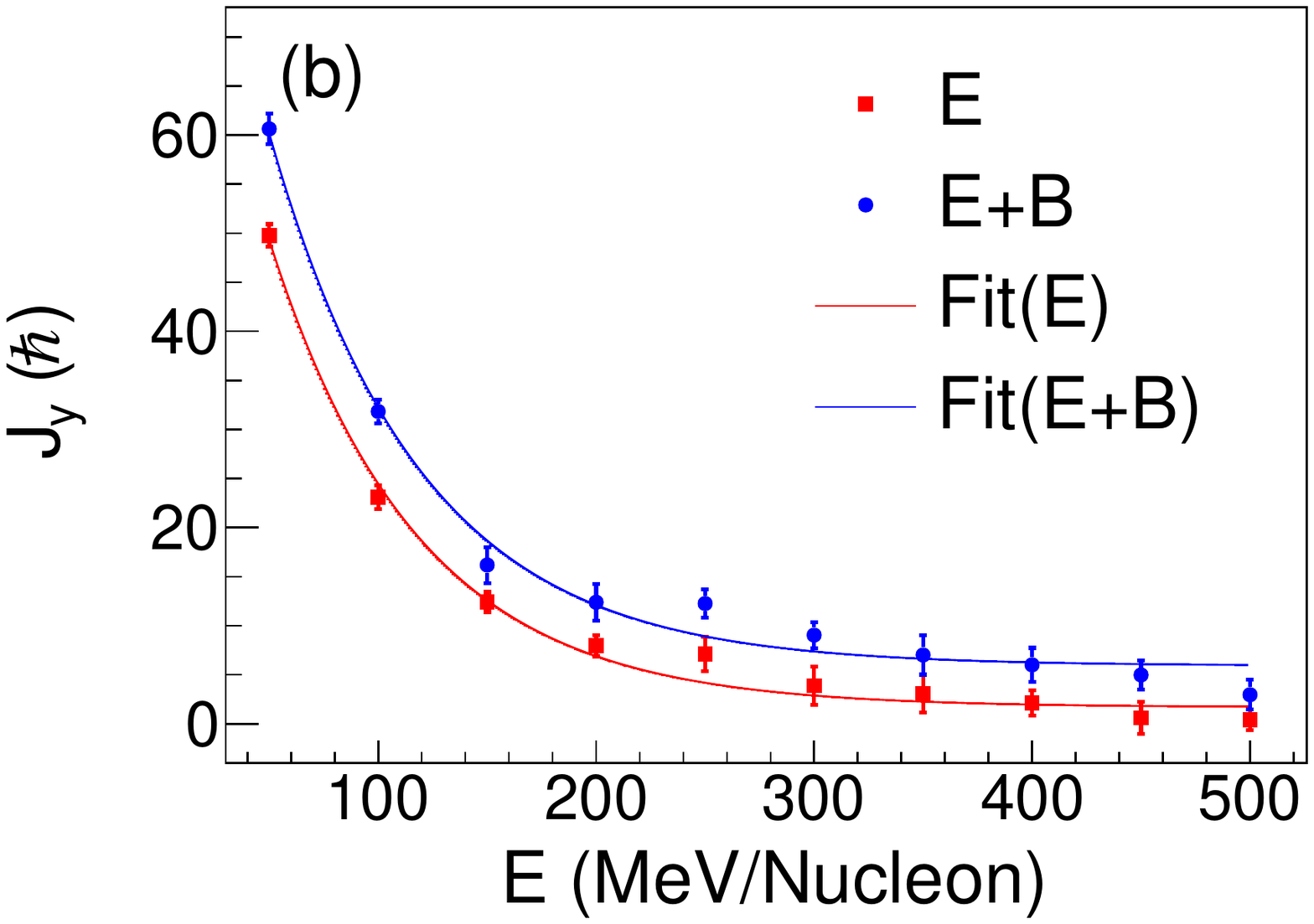}
\caption{
Temperature (a) and the $y$-component of angular momentum (b) as a function of incident energy for $^{40}$Ca excited by $^{16}$O in cases of without and with magnetic field.}
\label{fig:ImpactBT}
\end{figure}

By Gaussian fitting to the GDR spectrum, the peak energy $E_{\gamma}^{c}$, the strength $(dP/dE_{\gamma})^{c}$, and the width $\Gamma_{\gamma}^{c}$ can be quantitatively extracted. In Eq.~(\ref{Magneticfield01}), particle acceleration is considered. One could notice that accelerated charged particles emit EM waves and lose their energies. As shown in Fig.~\ref{fig:ImpactA}, one can see that when particle acceleration is taken into account, energy position and strength of the GDR  decrease and the width of the GR becomes wider. However, such effects are small. So we neglect particle acceleration effect in the following calculations.

Fig.~\ref{fig:ImpactMagneticfield} (a) shows the comparison between our calculations and the experimental data from Ref.~\cite{Ahrens75,Erokhova03}. The value of $E_{\gamma}^{c}$ is fitted to the data around 100 MeV/nucleon. As incident energy increases, $E_{\gamma}^{c}$ increases and is higher than the data when E $>$ 200 MeV/nucleon. And one can notice that for all of peak energy, strength and width of GDR spectrum, they are enhanced when the magnetic field is switched on.
In addition, one can see that the width decreases with  the increasing of incident energy when the incident energy is less than 200 MeV/nucleon but keeps stable as incident energy further increases. Also the gap is different between the cases for more and for less than 250 MeV/nucleon. Such a different trend could be related to different effects of the magnetic field generated by the projectile on the target.

\subsection{Impacts on temperature and angular momentum of $^{40}$Ca}
\label{ResultsB}

In Fig.~\ref{fig:ImpactMagneticfield}, we find that the magnetic field enhances the peak energy, strength and width of GDR spectrum. As we know, here the GDR mode is based on the excited state of the nucleus. Therefore here we check the excitation energy of target nucleus $^{40}$Ca.
Fig.~\ref{fig:ImpactBT}(a) displays the incident energy dependence of temperature which is obtained by $E^{\ast}$ =
$\it{k} T^2$, where $E^{\ast}$ = $(E_{total}-E_{ground})/A$ is the calculated excitation energy per nucleon ~\cite{Ma-1997}. The level density parameter $k$ is taken as $1/8$ and $A$ represents the mass number of $^{40}$Ca \cite{Reisdorf81}.
Since we are just focusing on ultra-peripheral collision, the higher the incident energy is, the shorter the interaction time between $^{40}$Ca and $^{16}$O is, and then the lower the excitation energy of $^{40}$Ca is. Thus, the $T$ of $^{40}$Ca decreases with the increasing of incident energy.
However, the influence of magnetic field on the temperature differs from the results of GDR width  as shown in Fig.~\ref{fig:ImpactMagneticfield}(c).
It indicates that something affected by the magnetic field other than temperature, such as
rotation of nucleus, caused the different trend of the width~\cite{Alhassid99}. 
Thus we check the angular momentum of nucleus, which can be calculated by $\boldsymbol{J}$ = $\sum_{i}\boldsymbol{r_{i}}\times\boldsymbol{p_{i}}$ where $i$ represents each nucleon \cite{Ma95,XuZW} under the center of mass system of $^{40}$Ca,  and obtain the results as shown in Fig.~\ref{fig:ImpactBT}(b).
One can see that the dependence of $y$-component of angular momentum $J_{y}$ on incident energy with magnetic field is higher than that one without magnetic field. As in Refs.~\cite{Alhassid99,DS20}, the width of GDR increases with angular momentum of nucleus. So here the enhancement of $J_{y}$ by magnetic field could be related to the results in Fig.~\ref{fig:ImpactMagneticfield}(c). Also, combining Fig.~\ref{fig:ImpactBT}(a) and Fig.~\ref{fig:ImpactBT}(b), we found that the magnetic field reduces $T$ and increases $J_{y}$ below 250 MeV/nucleon. While above 250 MeV/nucleon, the magnetic field increases both of $T$ and $J_{y}$ that could explain why the enhancement of width of GDR becomes larger above 250 MeV/nucleon as shown in Fig.~\ref{fig:ImpactMagneticfield}.
One can also notice that for the low beam energy, the angular momentum is large even without magnetic field. So in low energy region, it may be not only vibrational mode for GDR but also mixed modes due to rotation  that determine the width of the GDR.  In this case, it is hard  to extract a pure vibrational mode for GDR. However, as beam energy increases, the angular momentum becomes small which would reduce such a mixed mode effect. Thus it could be the reason for the fact that  at higher beam energy region, the energy position of the ``GDR" is close to the experimental result. Note that the experimental data corresponds to the case with only the electric field.

\begin{figure}[htb]
\setlength{\abovecaptionskip}{0pt}
\setlength{\belowcaptionskip}{0pt}
\centering\includegraphics[scale=0.58]{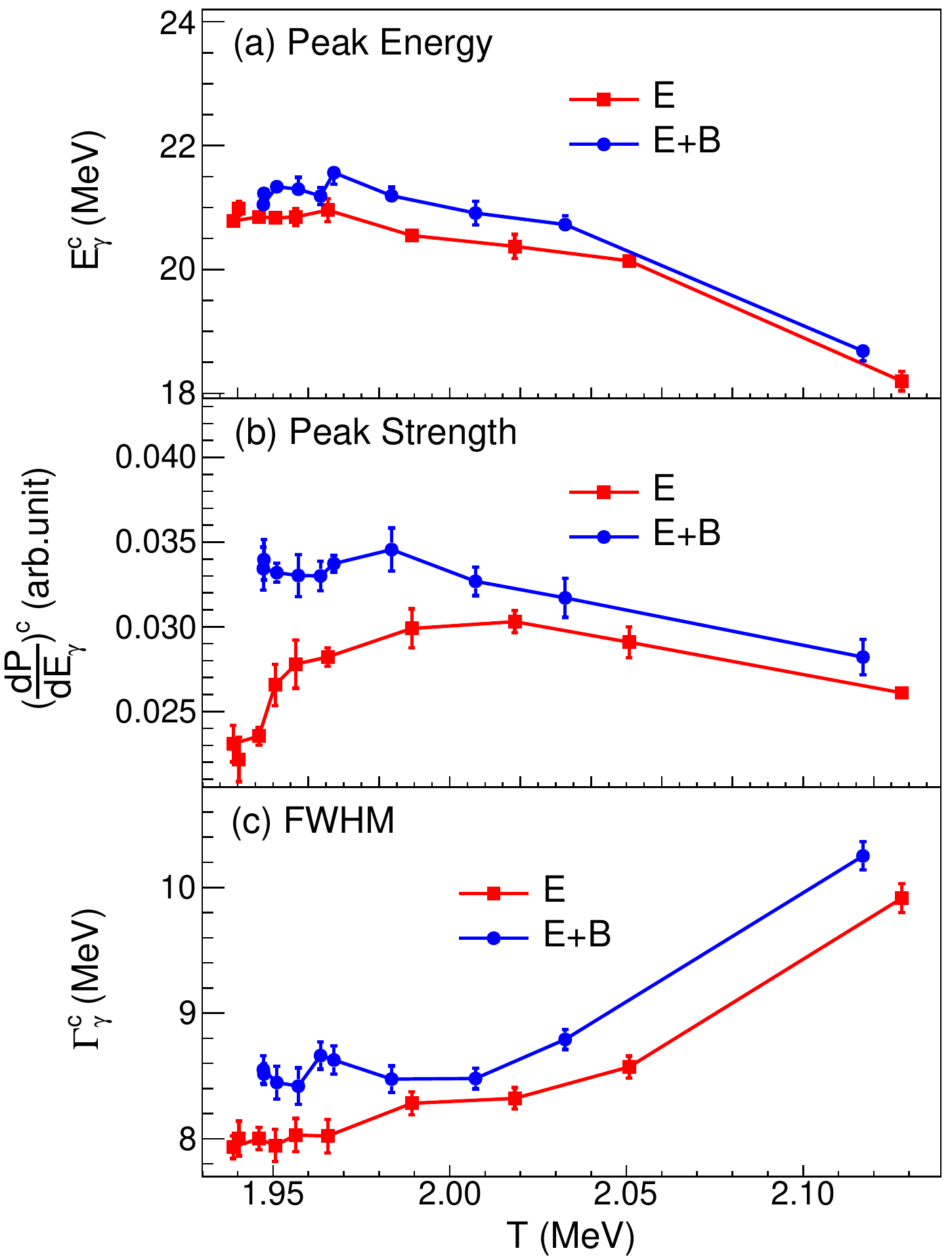}
\caption{The dependences of peak energy (a), strength (b) and width of GDR spectrum (c)  on temperature for $^{40}$Ca in the cases  without or with magnetic field.}
\label{fig:Impact3paraT}
\end{figure}

\subsection{Correlation between the GDR and temperature}
\label{ResultsC}

Considering that the GDR with finite temperature is an interesting issue~\cite{DS20}, we plot temperature dependence of GDR variables in  Fig.~\ref{fig:Impact3paraT}.
It is clear to see that for $E_{\gamma}^{c}$, it decreases as the temperature increases. Actually, in analogy with a classical oscillator with a friction force, the higher the temperature, the greater the probability of two-body dissipation, which means that the frequency of collective motion is reduced~\cite{Toro99,GuoCQ}, thus the $E_{\gamma}^{c}$ is decreased. Certainly, for $\Gamma_{\gamma}^{c}$, it increases with  temperature, which is consistent with the results in Refs.~\cite{NDD05,Mondal18}. Additionally, the enhancement of two-body dissipation reduces the available phase space of the $\Gamma_{\gamma}^{c}$ resulting in the decrease of the $(dP/dE_{\gamma})^{c}$ at higher $T$~\cite{Srijit21}. For the magnetic field dependence at different temperatures, we can refer to Fig.~\ref{fig:ImpactBT}(a).  Moreover, the magnetic field does not only affect temperature but also angular momentum of nucleus as shown in Fig.~\ref{fig:ImpactBT}(b). Thus the magnetic effect on GDR indicates a mechanism of mixture of temperature and angular momentum effects.

\section{Conclusion}
\label{summary}
In summary, by taking into account the magnetic field in the extended quantum molecular dynamics model (EQMD), we analyzed its effects on GDR of $^{40}$Ca through  the Coulomb  excitation process of $^{16}$O + $^{40}$Ca collision. We found that the magnetic field enhances the peak energy, strength and width of GDR spectrum. By checking the effects on temperature and angular momentum, we found that magnetic field increases the angular momentum of $^{40}$Ca. Thus how the magnetic field acts on angular momentum and nucleus shape evolution is an interesting issue and should be explored in future study. As for the correlation between the GDR and temperature for excited $^{40}$Ca under electric and magnetic fields, the peak energy and strength decreases with the increasing of temperature, but the width enhances,  which is in overall agreement with other theoretical work. The current work sheds light on  resurveying the magnetic field effects on nuclear collective excitation in low-intermediate energy heavy-ion collisions.

\vspace{0.5cm}
This work was supported in part by the National Natural Science Foundation of China under contract Nos.  11890710, 11890714, and 12147101, and the Guangdong Major Project of Basic and Applied Basic Research No. 2020B0301030008.

\end{CJK*}
\end{document}